\documentclass[runningheads,a4paper]{llncs}

\usepackage{amssymb}
\usepackage{graphicx}
\usepackage{enumerate}
\usepackage{paralist}
\usepackage{hyperref}
\usepackage{url}

\usepackage[ruled,section]{algorithm}
\usepackage[noend]{algorithmic}
\usepackage{amssymb}
\usepackage{fca}
\usepackage{amsmath}
\usepackage{amsfonts}
\usepackage{alltt}

\usepackage{url}
\usepackage{amsmath}
\urldef{\mailsa}\path|dignatov@hse.ru|
\newcommand{\keywords}[1]{\par\addvspace\baselineskip
\noindent\keywordname\enspace\ignorespaces#1}

\newcommand{\K}{\mathbb{K}}

\begin{document}

\mainmatter  

\setcounter{page}{57}

\titlerunning{An FCA-based Boolean Matrix Factorisation for Collaborative Filtering}
\authorrunning{Elena Nenova \and Dmitry I. Ignatov \and Andrey V. Konstantinov}

\title{An FCA-based Boolean Matrix Factorisation for Collaborative Filtering}

\titlerunning{An FCA-based Boolean Matrix Factorisation for Collaborative Filtering}

%
%
\author{Elena Nenova\inst{2,1} \and Dmitry I. Ignatov\inst{1} \and Andrey V. Konstantinov\inst{1}}
\authorrunning{}

\institute{National Research University Higher School of Economics, Moscow\\
\mailsa\\
\url{http://www.hse.ru}
\and
Imhonet, Moscow\\
\url{http://imhonet.ru}
}

%
%

\toctitle{An FCA-based Boolean Matrix Factorisation for Collaborative Filtering}
\tocauthor{}
\maketitle

\begin{abstract}
We propose a new approach for Collaborative filtering which is based on Boolean Matrix Factorisation (BMF) and Formal Concept Analysis. In a series of experiments on real data (Movielens dataset) we compare the approach with the SVD- and NMF-based algorithms in terms of Mean Average Error (MAE). One of the experimental consequences is that it is enough to have a binary-scaled rating data to obtain almost the same quality in terms of MAE by BMF than for the SVD-based algorithm in case of non-scaled data.

\end{abstract}

\section{Introduction}

Recently Recommender Systems is one of the most popular subareas of Machine Learning.  In fact, the recommender algorithms based on matrix factorisation techniques (MF) has become industry standard.

Among the most frequently used types of Matrix Factorisation we definitely should mention Singular Value Decomposition (SVD) \cite{svd} and its various modifications like Probabilistic Latent Semantic Analysis (PLSA) \cite{PLSA}. However, the existing similar techniques, for example, non-negative matrix factorisation (NMF) \cite{nmf1,nmf2,nmf3} and Boolean matrix factorisation (BMF) \cite{bmf}, seem to be less studied. An approach similar to matrix factorization is biclustering which was also successfully applied in recommender system domain \cite{Symeonidis:2008,Ignatov:2012b}. For example, Formal Concept Analysis \cite{1999:Ganter:FCA} can also be used as a biclustering technique and there are some of its applications in recommenders' algorithms \cite{Bridge:2006,Ignatov:2008}.

\underline{The aim of this paper} is to compare recommendation quality of some of the aforementioned techniques on real datasets and try to investigate the methods' interrelationship. It is especially interesting to conduct experiments on comparison of recommendations quality in case of an input matrix with numeric values and in case of a Boolean matrix in terms of Precision and Recall as well as MAE. Moreover, one of the useful properties of matrix factorisation is its ability to keep reliable recommendation quality even in case of dropping some insufficient factors. For BMF this issue is experimentally investigated in section \ref{exp}.

\underline{The novelty} of the paper is defined by the fact that it is  a first time when BMF based on Formal Concept Analysis \cite{1999:Ganter:FCA} is investigated in the context of Recommender Systems.

\underline{The practical significance} of the paper is determined by demands of the recommender systems' industry, that is to gain reliable quality in terms of Mean Average Error (MAE), Precision and Recall as well as time performance of the investigated method.

The rest of the paper consists of five sections. The second section is an introductory review of the existing MF-based recommender approaches. In the third section  we describe our recommender algorithm which is based on Boolean matrix factorisation using closed sets of users and items (that is FCA). Section 4 contains methodology of our experiments and results of experimental comparison of different MF-based recommender algorithms by means of cross-validation in terms of MAE, Precision and Recall. The last section concludes the paper.

\section{Introductory review of some matrix factorisation approaches}\label{review}


In this section we briefly describe different approaches to the  decomposition of  both real-valued and Boolean matrices. Among the methods of the SVD group we describe only SVD. We also discuss nonnegative matrix factorization (NMF) and Boolean matrix factorization (BMF).

\subsection{Singular Value Decomposition (SVD)}

Singular Value Decomposition (SVD) is a decomposition of a rectangular matrix $ A \in \mathbb {R} ^ {m \times n} (m> n) $ into the product of three matrices

\begin{equation} A=U\left(
       \begin{array}{c}
         \Sigma \\
         0 \\
       \end{array}
     \right)V^T,\end{equation}


where $ U \in \mathbb {R} ^ {m \times m} $ and $ V \in \mathbb {R} ^ {n \times n} $ are orthogonal matrices, and $ \Sigma \in \mathbb {R } ^ {n \times n} $ is a diagonal matrix such that $ \Sigma = diag (\sigma_1, \ldots, \sigma_n) $ and $ \sigma_1 \geq \sigma_2 \geq \ldots \geq \sigma_n \geq0 $. The columns of the matrix  $ U $ and $ V $ are called singular vectors, and the numbers $ \sigma_i $ are singular values ​\cite{svd}.


In the context of recommendation systems rows of $U$ and $V$  can be  interpreted  as vectors of the user's  and items's  loyalty (attitude) to a certain topic (factor), and the corresponding singular values as the importance of the topic among the others. The main disadvantage is in the fact that the matrix may contain both positive and negative numbers; the last ones are difficult to interpret.


The advantage of SVD for recommendation systems is that this method allows to obtain the vector of its loyalty to certain topics for a new user without SVD decomposition of the whole matrix.


The evaluation of computational complexity of SVD according to \cite{bau:1997} is $O(mn^2)$  floating-point operations if $m \geq n$  or more precisely  $2mn^2 + 2n^3$.


Consider as an example the following table of  movie ratings:

\begin{table}[ht]
\caption{Movie rates}\label{tbl-data}
\begin{center}
\begin{small}
            \begin{tabular}{|c|c|c|c|c|c|c|c|c|}
              \hline
               & The Artist  &Ghost & Casablanca &Mamma Mia!&  Dogma& Die Hard &Leon   \\
           \hline
            User1& 4 & 4 & 5 & 0 & 0 & 0 & 0 \\
            \hline
            User2 &5 & 5 & 3 & 4 & 3 & 0 & 0 \\
            \hline
            User3 &0 & 0 & 0 & 4 & 4 & 0 & 0 \\
            \hline
            User4 &0 & 0 & 0 & 5& 4 & 5 & 3 \\
             \hline
             User5&0 & 0 & 0 & 0 & 0 & 5 &  5\\
             \hline
            User6&0 & 0 & 0 & 0 & 0 & 4 & 4 \\
              \hline
            \end{tabular}
\end{small}
\end{center}
\end{table}


This table corresponds to the following matrix of ratings:
 $$A=\left(
    \begin{array}{ccccccc}
         4 & 4 & 5 & 0 & 0 & 0 & 0 \\
        5 & 5 & 3 & 4 & 3 & 0 & 0 \\
        0 & 0 & 0 & 4 & 4 & 0 & 0 \\
        0 & 0 & 0 & 5& 4 & 5 & 3 \\
        0 & 0 & 0 & 0 & 0 & 5 &  5\\
        0 & 0 & 0 & 0 & 0 & 4 & 4 \\
    \end{array}
  \right)
.$$


From the  SVD  matrix decomposition we get:
$$U=\left(
      \begin{array}{cccccc}
    0.31  &   0.48 &   -0.49  &  -0.64  &  -0.06  &   0\\
    0.58  &   0.50 &    0.03 &    0.63  &   0.06 &   0\\
    0.29 &    0  &   0.57 &   -0.23 &   -0.72  &   0\\
    0.57  &  -0.37 &    0.31  &  -0.30  &   0.57 &    0\\
    0.29 &   -0.47 &   -0.43  &   0.15  &  -0.28 &   -0.62\\
    0.23 &   -0.37 &   -0.35  &   0.12 &   -0.22 &   0.78\\
      \end{array}
    \right),
    $$
    $$
    \left(
       \begin{array}{c}
         \Sigma \\
         0 \\
       \end{array}
     \right)= \left(
      \begin{array}{ccccccc}
       12.62    &  0     &    0    &     0    &     0     &    0  &       0\\
         0  & 10.66     &    0   &      0     &    0   &      0     &    0\\
         0   &      0   & 7.29   &      0     &    0    &     0     &    0\\
         0    &     0   &      0  &  1.64     &    0    &     0    &     0\\
         0    &     0   &      0    &     0   & 0.95    &     0    &     0\\
         0     &    0   &      0     &    0   &      0  &  0  &      0\\
      \end{array}
    \right),$$
    $$
      V^T=   \left(
      \begin{array}{ccccccc}
        0.32  &  0.41 &  -0.24  &  0.36  &  0.07 &   0.70 &  0.13\\
        0.32  &  0.41&   -0.24  &  0.36 &   0.07 &  -0.62 &  -0.35\\
        0.26  &  0.37  & -0.32 &  -0.79  & -0.12 &  -0.06  & 0.17\\
        0.50  &  0.01 &   0.55  &  0.05  &  0.24 &  -0.21 &   0.57\\
        0.41 &   0.01 &   0.50 &  -0.14  & -0.42  & 0.21  & -0.57\\
        0.42 &  -0.53 &  -0.27  & -0.15  &  0.57 &   0.10 &  -0.28\\
        0.33 &  -0.46 &  -0.36  &  0.21  & -0.63  & -0.10  &  0.28\\
      \end{array}
    \right).$$


It can be seen that the greatest weight have the first three singular values, which is confirmed by the calculations:

$$\frac{\sum\limits_{i=1}^{3}\sigma_i^2}{\sum\sigma_i^2}\cdot 100\% \approx 99\%.$$

\subsection{Non-negative matrix factorisation (NMF)}

Non-negative Matrix Factorization (NMF) is a decomposition of non-negative matrix $ V \in \mathbb {R} ^ {n \times m} $  for a given number  $ k $  into  the product of two non-negative matrices $ W \in\mathbb{R}^{n\times k}$ and $H\in \mathbb{R}^{k\times m}$ such that

\begin{equation}
V\approx WH.
\end{equation}


NMF is widely used in such areas as finding the basis vectors for images, discovering molecular structures, etc. \cite{nmf1}.

Consider the following matrix of ratings:
 $$V=\left(
    \begin{array}{ccccccc}
         4 & 4 & 5 & 0 & 0 & 0 & 0 \\
        5 & 5 & 3 & 4 & 3 & 0 & 0 \\
        0 & 0 & 0 & 4 & 4 & 0 & 0 \\
        0 & 0 & 0 & 5& 4 & 5 & 3 \\
        0 & 0 & 0 & 0 & 0 & 5 &  5\\
        0 & 0 & 0 & 0 & 0 & 4 & 4 \\
    \end{array}
  \right).
$$


Its decomposition into the product of two non-negative matrices for $k = 3$ can be, for example, like this:
$$V=\left(
    \begin{array}{ccc}
        2.34  &  0&        0\\
    2.32  & 1.11  &       0\\
         0  &  1.28   &      0\\
         0  &  1.46 &   1.23\\
         0  &  0  & 1.60\\
         0   & 0  &  1.28\\
    \end{array}
  \right) \cdot \left( \begin{array}{ccccccc}
        1.89  &  1.89  &  1.71   & 0.06 &  0 &   0 &        0\\
    0.13  &  0.13 &   0  &  3.31  &  2.84  &  0.27 &  0\\
         0  &       0    &     0  &  0.03  &  0   & 3.27  &  2.93\\
    \end{array}\right).$$

\subsection{Boolean Matrix Factorisation (BMF) based on Formal Concept Analysis (FCA)}

\subsubsection{Basic FCA definitions. }


Formal Concept Analysis (FCA) is a branch of applied mathematics and it studies (formal) concepts and their hierarchy. The adjective ``formal'' indicates a strict mathematical definition of a pair of sets, called, the extent and intent. This formalisation is possible because the use of the algebraic lattice theory.


\textsc{Definition 1.}  {\em Formal context} $ \ K $ is a triple $ (G, M, I) $, where $ G $ is the set of {\em objects}, $ M $ is the set of {\em attributes }, $ I \subseteq G \times M $ is a binary relation.


The binary relation $ I $ is interpreted as follows: for $ g \in G $, $ m \in M ​​$ we write $ gIm $ if the object $ g $  has the attribute $ m $.

For a formal context $ \K = (G, M, I) $ and any $ A \subseteq G $ and $ B \subseteq M $  a pair of mappings is defined:

$$A^{\prime}= \{m\in M\mid gIm \ \mbox{ for all } g\in A\},$$
$$ B^{\prime} = \{g\in G\mid gIm \ \mbox{ for all } m\in B\},$$
these mappings define Galois connection between partially ordered sets $ (2^G, \subseteq) $ and $ (2^M, \subseteq) $ on disjunctive union of  $G$ and  $M$.
The set  $A$ is called  {\em closed set}, if $A^{\prime\prime} =A$ \cite{1989:Birkhoff:TLrus}.


\textsc{Definition 2.}  A {\em formal concept} of the formal context  $\K = (G, M, I)$ is a pair  $(A, B)$, where $A\subseteq G$, $B\subseteq M$, $A^{\prime} = B$ and $B^{\prime} = A$. Set $A$ is called the {\em extent}, and  $B$  is the {\em intent} of the formal concept  $(A, B)$.


It is evident that extent and intent of any formal concept are closed sets.

The set of formal concepts of a context $\K$ is denoted by $ \BGMI$.

\subsubsection{Description of FCA-based BMF}


Boolean matrix factorization (BMF) is a decomposition of the original matrix   $ I\in \{0,1 \}^{n\times m} $, where $I_ {ij}\in\{0,1 \}, $  into a Boolean matrix product $ P \circ Q $ of binary matrices $ P \in \{0,1 \}^{n \times k} $ and $ Q \in \{0,1 \} ^ {k \times m} $ for the smallest possible number of $k. $
We define boolean matrix product as follows:
$$(P\circ Q)_{ij}=\bigvee_{l=1}^kP_{il}\cdot Q_{lj},$$

where $ \bigvee $ denotes disjunction, and $ \cdot $ conjunction.


Matrix $ I $ can be considered as a matrix of binary relations between set $ X $ objects (users), and the set $ Y $ attributes (items that users have evaluated). We assume that $ xIy $  iff  user $x$  estimated object $y$. The triple $ (X, Y, I) $ is clearly composes a formal context.


Consider the set $ \mathcal {F} \subseteq \mathcal {B} (X, Y, I) $, a subset of all formal concepts of context $ (X, Y, I) $, and introduce the matrices $ P_ {\mathcal {F}} $ and $ Q_ {\mathcal {F}}: $
$$(P_{\mathcal{F}})_{il}=\left\{
                           \begin{array}{ll}
                             1, i\in A_l,\\
                             0, i\notin A_l,
                           \end{array}
                         \right.
  \ \ (Q_{\mathcal{F}})_{lj}=\left\{
                               \begin{array}{ll}
                                 1, j\in B_l, \\
                                 0, j\notin B_l.
                               \end{array}
                             \right.
  $$
We can consider the decomposition of the matrix $I$ into binary matrix product $ P_ \mathcal {F} $ and $ Q_ \mathcal {F} $ as described above. The following theorems are proved in \cite{bmf}:
\begin{enumerate}
  \item[Theorem 1.] 
(Universality of formal concepts as factors). For every $I$ there is  $\mathcal{F}\subseteq \mathcal{B}(X,Y,I)$,  such that  $I=P_\mathcal{F}\circ Q_\mathcal{F}.$
  \item[Theorem 2.]
(Optimality of formal concepts as factors). Let $I=P\circ Q$  for $n\times k$ and $k\times m$ binary matrices $P$ and $Q$. Then there exists a $\mathcal{F}\subseteq \mathcal{B}(X,Y,I)$ of formal concepts of $I$  such that $|\mathcal{F}|\leq k$ and  for the $n \times |F|$ and $|F| \times m$ binary matrices $P_\mathcal{F}$ and $Q_\mathcal{F}$ we have $I=P_\mathcal{F}\circ Q_\mathcal{F}.$
\end{enumerate}
There are  several algorithms for finding $P_ \mathcal {F}$ and $Q_ \mathcal {F}$ by calculating formal concepts based on these theorems \cite{bmf}.


The algorithm we use (Algoritm 2 from \cite{bmf}) avoid the computation of all the possible formal concepts and therefore works much faster \cite{bmf}. Time estimation of the calculation algorithm in the worst case yields $O(k|G||M|^3) $, where $ k $ is the number of found factors, $|G|$ is the number of objects, $|M|$  this the number of attributes.


Transform the  matrix of ratings described above, to a boolean matrix, as follows:
 $$\left(
    \begin{array}{ccccccc}
         4 & 4 & 5 & 0 & 0 & 0 & 0 \\
        5 & 5 & 3 & 4 & 3 & 0 & 0 \\
        0 & 0 & 0 & 4 & 4 & 0 & 0 \\
        0 & 0 & 0 & 5& 4 & 5 & 3 \\
        0 & 0 & 0 & 0 & 0 & 5 &  5\\
        0 & 0 & 0 & 0 & 0 & 4 & 4 \\
    \end{array}
  \right)\Rightarrow\left(
    \begin{array}{ccccccc}
         1 & 1 & 1 & 0 & 0 & 0 & 0 \\
        1 & 1 & 1 & 1 & 1 & 0 & 0 \\
        0 & 0 & 0 & 1 & 1 & 0 & 0 \\
        0 & 0 & 0 & 1& 1 & 1 & 1 \\
        0 & 0 & 0 & 0 & 0 & 1 &  1\\
        0 & 0 & 0 & 0 & 0 & 1 & 1 \\
    \end{array}
  \right)=I.
$$
The decomposition of the matrix $ I $ into the Boolean product of $ I = A_ {\mathcal {F}} \circ B_ {\mathcal {F}} $ is the following:
$$\left(
    \begin{array}{ccccccc}
         1 & 1 & 1 & 0 & 0 & 0 & 0 \\
        1 & 1 & 1 & 1 & 1 & 0 & 0 \\
        0 & 0 & 0 & 1 & 1 & 0 & 0 \\
        0 & 0 & 0 & 1& 1 & 1 & 1 \\
        0 & 0 & 0 & 0 & 0 & 1 &  1\\
        0 & 0 & 0 & 0 & 0 & 1 & 1 \\
    \end{array}
  \right)= \left( \begin{array}{ccc}
         1 & 0 & 0 \\
        1 & 1 & 0 \\
        0 & 1 & 0 \\
        0 & 1& 1 \\
        0 & 0 & 1 \\
        0 & 0 & 1 \\
    \end{array}
  \right)
  \circ  \left( \begin{array}{ccccccc}
         1 & 1 & 1 & 0 & 0 & 0 & 0 \\
        0 & 0 & 0 & 1 & 1 & 0 & 0 \\
        0 & 0 & 0 & 0 & 0 & 1 &1 \\
          \end{array}
  \right).$$

This example shows that the algorithm has identified three factors that significantly reduces the dimensionality of the data.
\subsection{General scheme of user-based recommendations}


Once the matrix of rates is factorized we need to learn how to compute recommendations for users and to evaluate whether a particular method handles well with this task.


For factorized matrices the already well-known algorithm based on the similarity of users can be applied, where for finding  $ K $ nearest neighbors we use not the original matrix of ratings $ A \in \mathbb {R} ^ {m \times n} $, but the matrix $ U \in \mathbb {R} ^ {m \times f} $, where $ m $ is a number of users, and $ f $ is a number of factors. After selection of $ K $  users, which are the  most similar to a given user, based on  the factors that are peculiar to them, it is possible, based on collaborative filtering formulas to calculate the projected rates for a given user.

After formation of the recommendations the performance of the recommendation system can be estimated by measures such as Mean Absolute Error (MAE), Precision and Recall.




\section{A recommender algorithm using FCA-based BMF}\label{FCAalg}

\subsection{kNN-based algorithm}

Collaborative recommender systems try to predict the utility of items for a particular user based on the items previously rated by other users.

Denote $u(c,s)$  the utility of item $s$ for user $c$. $u(c,s)$  is estimated based on the utilities $u(c_i,s)$ assigned to item $s$ by those users $c_i\in C$  who are ``similar'' to user $c$. For example, in a movie recommendation application, in order to recommend movies to user $c$, the collaborative recommender system  finds the users that have similar tastes in movies with $c$  (rate the same movies similarly). Then, only the movies that are most liked by those similar users  would be recommended.

Memory-based recommendation system, which are  based on the previous history of the ratings, are one of the key classes of collaborative recommendation systems.


Memory-based algorithms make rating predictions based on the entire collection of previously rated items by the users. That is, the value of the unknown rating $r_{c,s}$ for user $c$ and item $s$ is usually computed as an aggregate of the ratings of some other (usually, the $K$ most similar) users for the same item $s$:

$$r_{c,s}=aggr_{c'\in \widehat{C}}r_{c,s},$$
where $\widehat{C}$ denotes the set of $K$ users that are the most similar to user $c$ , who have rated item $s$. For example, function $aggr$ may has the following form \cite{2}
$$r_{c,s}=k \sum\limits_{c'\in \widehat{C}} sim(c',c)\times r_{c',s},\label{two}$$
where  $k$  serves as a normalizing factor and  selected as $k=1/\sum\limits_{c'\in \widehat{C}} sim(c,c')$.


The similarity measure between users $c$ and $c'$, $sim(c,c')$, is essentially a distance measure and is used as a weight, i.e., the more similar users $c$ and $c'$ are, the more weight rating $r_{c',s}$  will carry in the prediction of $r_{c,s}.$


The similarity between two users is based on their ratings of items that both users have rated. The two most popular approaches are {\em correlation} and {\em cosine-based}. One common strategy is to calculate all user similarities $sim(x,y)$  in advance and recalculate them only once in a while (since the network of peers usually does not change dramatically in a short time). Then, whenever the user asks for a recommendation, the ratings can be calculated on demand using precomputed similarities.

To apply this approach in case of FCA-based BMF recommender algorithm we simply consider as an input the user-factor matrices obtained after factorisation of the initial data.

\subsection{Scaling}\label{shkal}

In order to move from a matrix of ratings to a Boolean matrix, and use the results of Boolean matrix factorization, scaling is required. It is well known that scaling is a matter of expert interpretation of original data. In this paper, we use several variants of scaling and compare the results in terms of MAE.

\begin{enumerate}
\item $I_{ij}=1$ if $R_{ij} > 0$, else $I_{ij}=0$ (user $i$ rates item $j$).
\item $I_{ij}=1$ if $R_{ij} > 1,$ else $I_{ij}=0.$
\item $I_{ij}=1$ if $R_{ij }> 2,$ else $I_{ij}=0$.
\item $I_{ij}=1$ if $R_{ij} > 3$, else $I_{ij}=0.$
\end{enumerate}

\section{Experiments}\label{exp}

To test our hypotheses and study the behavior of recommendations based on the factorization of a  ratings matrix  by different methods we used MovieLens data. We used the part of data, containing 100,000 ratings, while considered only users who have given over 20 ratings.


User ratings are split into two sets, a training set consisting of 80 000 ratings, and   test set  consisting of 20 000 ratings. Original data matrix is $ 943 \times $ 1682, where the number of rows is the number of users and the number of columns is the number of rated movies (each film has at least one vote).

\subsection{The number of factors that cover $p$\% of evaluations in an input data for SVD and BMF}
The main purpose of matrix factorization  is a reduction of matrices dimensionality. Therefore we examine how the number of factors varies depending on the method of factorization, and depending on $ p $ \%  of the data that is covered by factorization.
For BMF the coverage of a matrix is calculated as the ratio of the number of ratings covered by Boolean factorization to the total number of ratings.
\begin{equation}\frac{|covered\_ratings|}{|all\_ratings|}\cdot 100\% \approx p_{BMF}\%,\end{equation}
 For SVD we use the following formula:
\begin{equation}\frac{\sum\limits_{i=1}^{K}\sigma_i^2}{\sum\sigma_i^2}\cdot 100\% \approx p_{SVD}\%,\end{equation}
where $K$ is the number of factors selected.

\begin{table}[ht]
\caption{Number of factors for SVD and BMF at different coverage level}\label{tbl-numFcov}
\begin{center}

\begin{tabular}{|c|c|c|c|}
  \hline
  p\% &  100\%&  80\%&60\%  \\ \hline
   SVD & 943 &175 &67  \\ \hline
   BMF & 1302 &402 &  223\\
     \hline
\end{tabular}

\end{center}
\end{table}

\subsection{MAE-based recommender quality comparison of SVD and BMF for various levels of evaluations coverage}\label{svd-bmf}


The main purpose of matrix factorisation  is a reduction of matrices dimensionality. As a result some part of  the original data remains uncovered, so it was interesting to explore how the quality of recommendations changes  based on  different factorisations, depending on the proportion of the data covered by factors.


Two methods of matrix factorisation were considered: BMF and SVD. The fraction of data covered by factors for SVD was calculated as $$ p \% = \frac {\sum \limits_ {i = 1}^{K} \sigma_i^2} {\sum \sigma_i^2} \cdot 100 \% , $$ and for BMF as $$ p \% = \frac{|covered\_ratings|} {|all \_ratings|} \cdot 100 \%. $$ To quality assessment we chose $ MAE $.

\begin{figure}
 \centering
\includegraphics[width=4.5in]{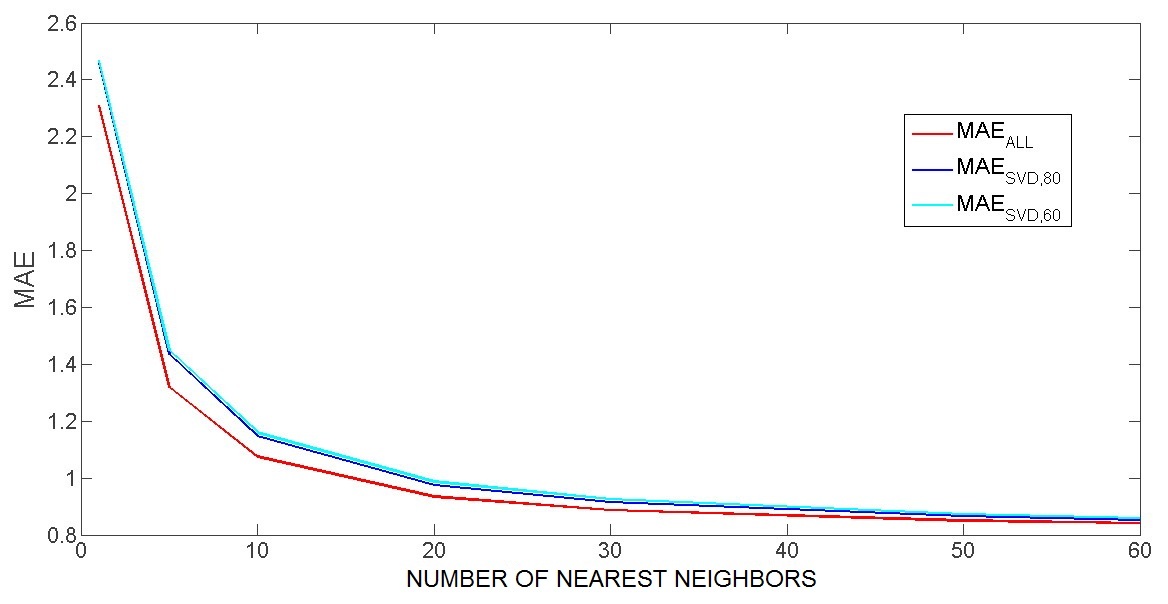}\\
  \caption{MAE dependence on the percentage of the data covered by SVD-decomposition, and the number of nearest neighbors.}
  \label{svd}
\end{figure}

\begin{figure}
 \centering
\includegraphics[width=4.55in]{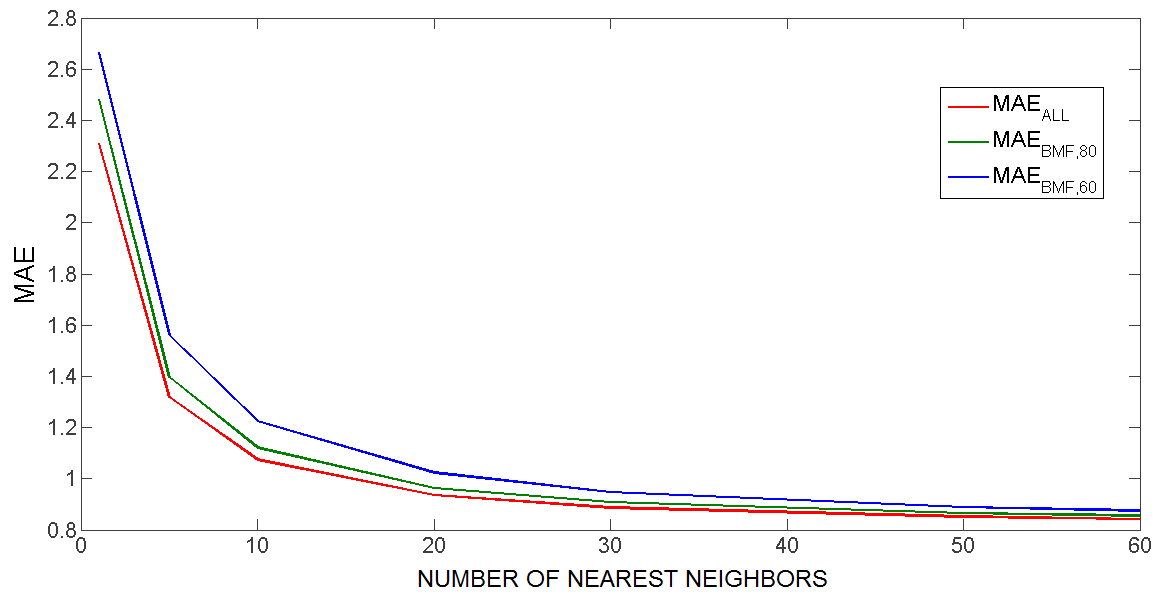}\\
  \caption{MAE dependence on the percentage of the data covered by BMF-decomposition, and the number of nearest neighbors.}
  \label{bmf}
\end{figure}

Fig.~\ref{svd} shows that $MAE_{SVD60} $, calculated for the model based on $ 60\% $ of factors, is not very different from $MAE_ {SVD80}$, calculated for the model built for $80\%$ factors. At the same time, for the recommendations based on a Boolean factorization covering $60\%$ and $ 80\% $ of the data respectively, it is clear that increasing the number of factors improves MAE, as shown in Fig.~\ref{bmf}.

\begin{table}[ht]
\caption{MAE for SVD and BMF at  80\% coverage level}\label{tbl-MAEcov}
\begin{center}

\begin{small}
\begin{tabular} {|c|c|c|c|c|c|c|c|}
  \hline
   Number of  neighbors & 1 & 5 & 10 & 20 & 30 & 50 & 60\\ \hline
  $MAE_{SVD80}$ & 2,4604	& 1.4355 &	1.1479	& 0.9750 &	0.9148	& 0.8652 &	0.8534 \\ \hline
   $MAE_{BMF80}$ & 	2.4813 &	1.3960	& 1.1215 &	0.9624 &	0.9093	& 0.8650 &	0.8552\\ \hline
    $MAE_{all} $& 2.3091 &	1.3185	& 1.0744  &	0.9350 &	0.8864 &	0.8509	& 0.8410\\  \hline
\end{tabular}
\end{small}

\end{center}
\end{table}


Table \ref{tbl-MAEcov} shows that the MAE for recommendations built on a Boolean factorisation covering 80 \% of the data for the number of neighbors less than 50 is better than the MAE for recommendations built on SVD factorization. It is also easy to see that $ MAE_{SVD80} $ and $ MAE_{BMF80} $ are different from $ MAE_{all} $ in no more than $1-7 \%$.

\subsection{Comparison of kNN-based approach and BMF by Precision and Recall}


Besides comparison of algorithms using MAE other evaluation metrics can also be exploited, for example

$$Recall=\frac{|objects\_in\_recommendation\cap objects\_in\_test|}{|objects\_in\_test|},$$

$$Precision=\frac{|objects\_in\_recommendation\cap objects\_in\_test|}{|objects\_in\_recommendation|}$$

and

$$F1=\frac{2\cdot Recall\cdot Precision}{Recall+Precision}.$$

It is widely spread belief that the larger Recall, Precision  and F1 are, the better is recommendation algorithm.

\begin{figure}
 \centering
\includegraphics[width=5in]{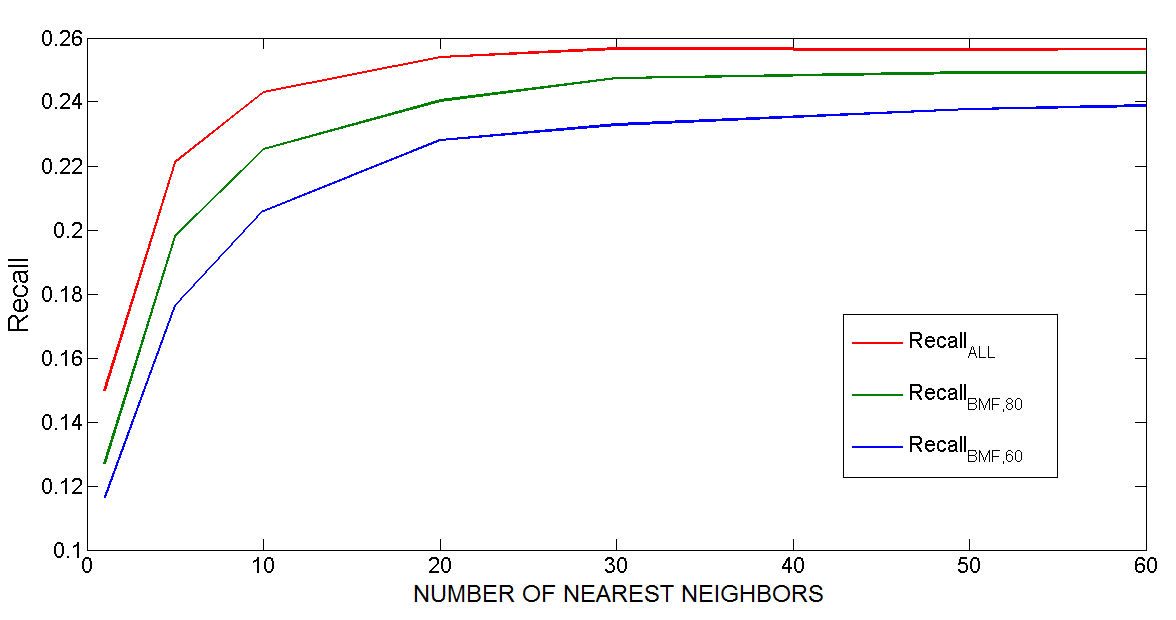}\\
  \caption{\textit{Recall  dependence on the percentage of data covered by BMF-decomposition, and the number of nearest neighbors.}}
  \label{recall}
\end{figure}
\begin{figure}[ht]
 \centering
\includegraphics[width=5in]{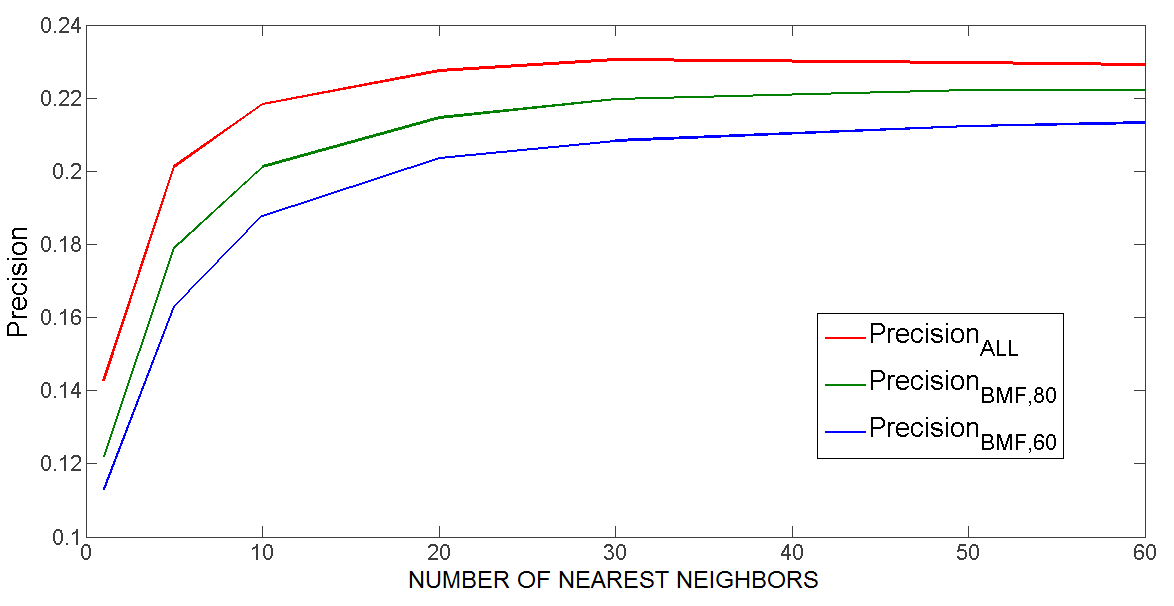}\\
  \caption{\textit{Precision  dependence on the percentage of data covered by BMF-decomposition, and the number of nearest neighbors.}}
  \label{precision}
\end{figure}
\begin{figure}[ht]
 \centering
\includegraphics[width=5in]{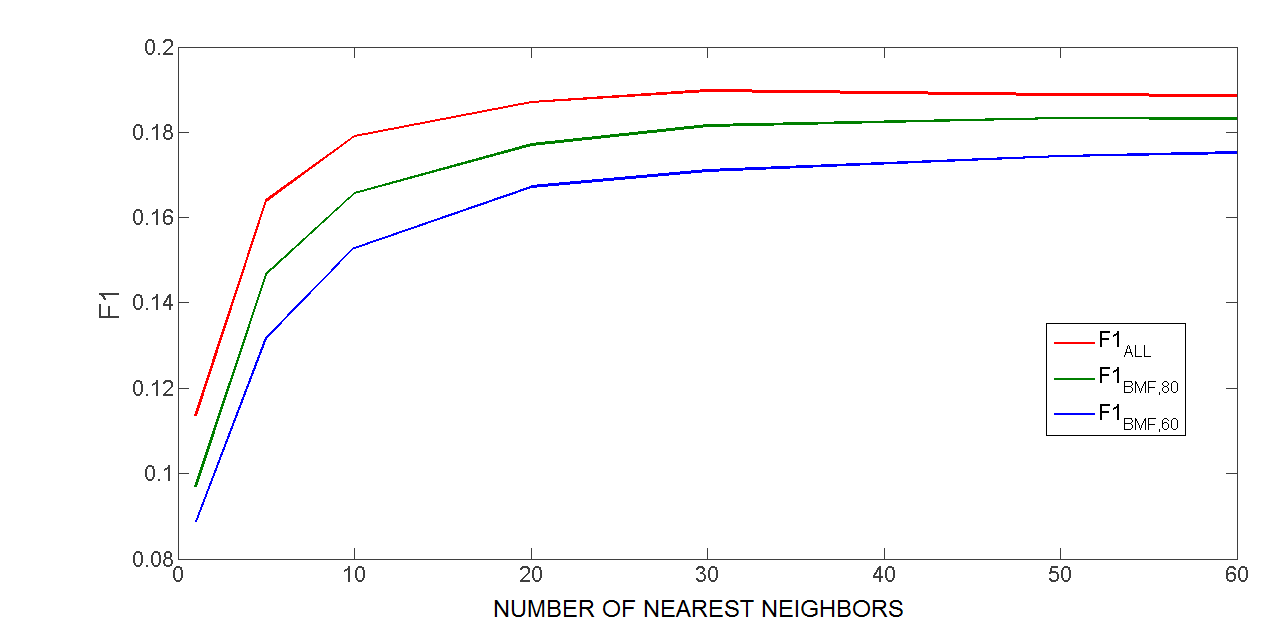}\\
  \caption{\textit{F1  dependence on the percentage of data covered by BMF-decomposition, and the number of nearest neighbors.}}
  \label{F1}
\end{figure}

Figures \ref{recall}, \ref{precision} and \ref{F1} show the dependence of relevant evaluation metrics on the percentage of the data covered by BMF-decomposition, and the number of nearest neighbors. The number of objects to recommend was chosen to be 20. The figures show that the recommendation based on the Boolean decomposition, is worse than recommendations built on the full matrix of ratings.

\subsection{Scaling influence on the recommendations quality for BMF in terms of MAE}
Another thing that was interesting to examine was the impact of scaling described in \ref{shkal} on the quality of recommendations. Four options of scaling were considered:

\begin{enumerate}
\item $I_{0,ij}=1$ if $A_{ij} > 0$, else $I_{ij}=0$ (user rates an item).
\item $I_{1,ij}=1$ if $A_{ij} > 1,$ else $I_{ij}=0.$
\item $I_{2,ij}=1$ if $A_{ij} > 2,$ else $I_{ij}=0$.
\item $I_{3,ij}=1$ if $A_{ij} > 3$, else $I_{ij}=0.$
\end{enumerate}

\begin{figure}[ht]
 \centering
\includegraphics[width=2.5in]{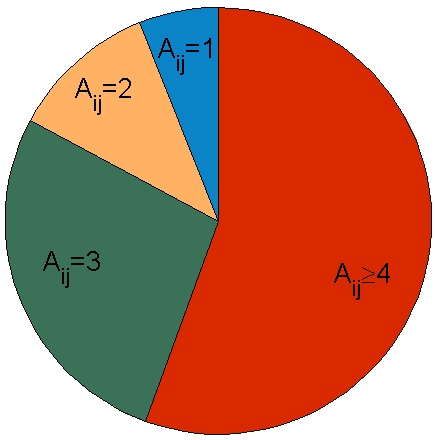}\\
  \caption{\textit{Ratings distribution in data.}}
  \label{circle}
\end{figure}

The distribution of ratings in data is on Figure~\ref{circle}

For each of the boolean matrices we calculate its Boolean factorisation, covering 60 \% and 80 \% of the data. Then recommendations are calculated just like in \ref{svd-bmf}.
It can be seen that for both types of data coverage $ MAE_1 $ is almost the same as $ MAE_0$, and $ MAE_ {2,3} $ is better than $ MAE_0$.

\begin{figure}[ht]
 \centering
\includegraphics[width=5in]{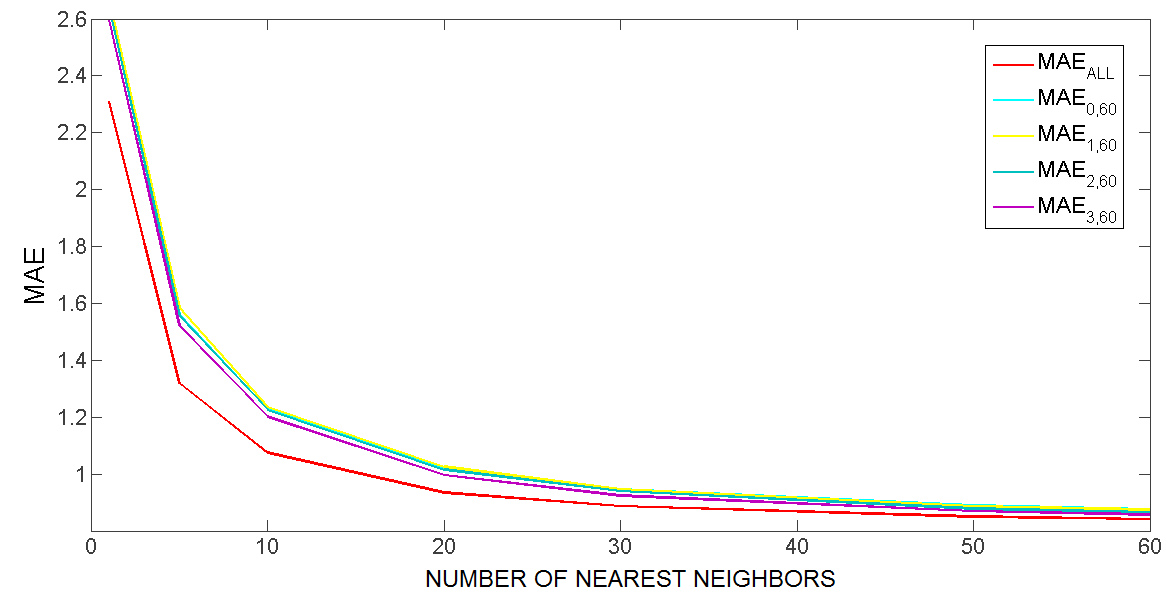}
\label{mae_shkal_60}
\caption{MAE dependance on  scaling and number of nearest  neighbors for 60\% coverage.}
\end{figure}

\begin{figure}[ht]
 \centering
\includegraphics[width=5in]{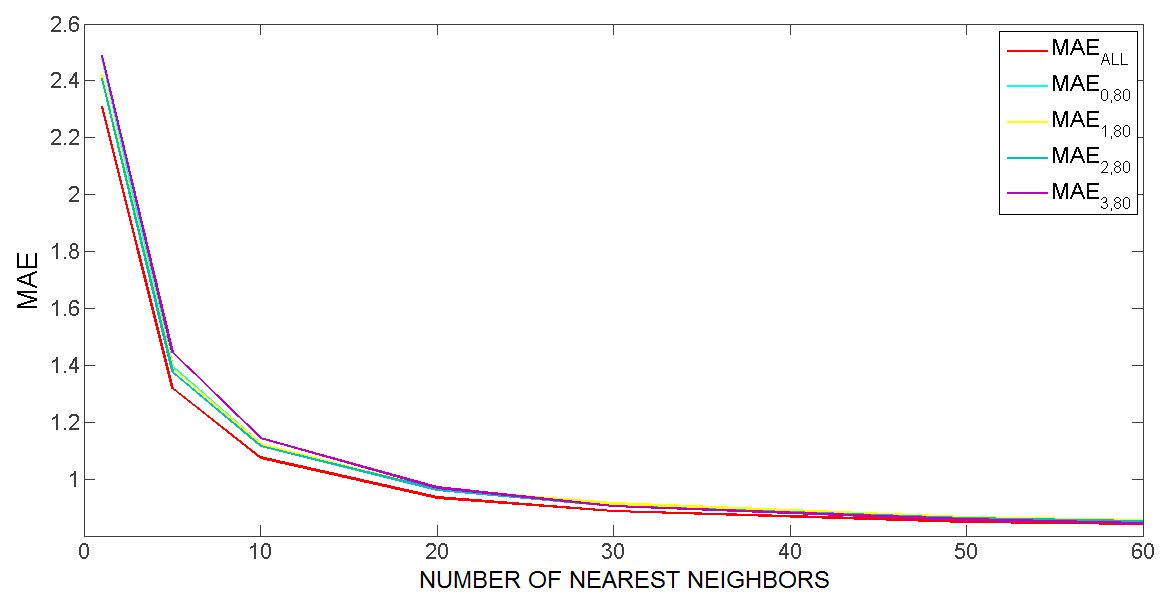}
\label{mae_shkal_80}
\caption{MAE depandance on  scaling and number of nearest  neighbors for 80\% coverage.}
\end{figure}

\subsection{Influence of data filtering on MAE for BMF kNN-based approach}

Besides the ability to search for $ K $ nearest neighbors not in the full matrix of ratings $ A \in \mathbb{R}^{n \times m} $, but in  the matrix $ U \in \mathbb{R} ^ {m \times f} $, where $ m $ is a number of users, and $ f $ is a number of factors, Boolean matrix factorization can be used to data filtering. Because the algorithm returns as an output not only matrices users-factors and factors-objects, but also the ratings that were not used for factoring, we can try to search for users, similar to the user, on the matrix consisting  only of ratings used for the factorization.

\begin{figure}[ht]
 \centering
\includegraphics[width=4.7in]{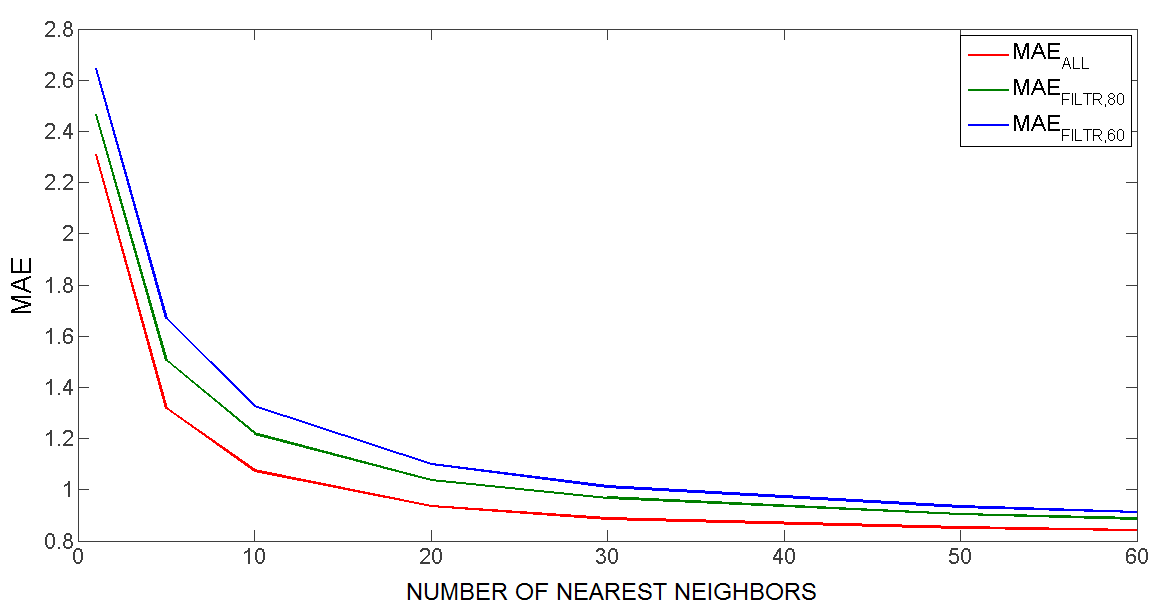}\\
  \caption{\textit{ MAE dependance on percentage of covered with filtration data and the  number of nearest  neighbors.}}
  \label{mae_filtr_all}
\end{figure}

\begin{figure}[ht]
 \centering
\includegraphics[width=4.7in]{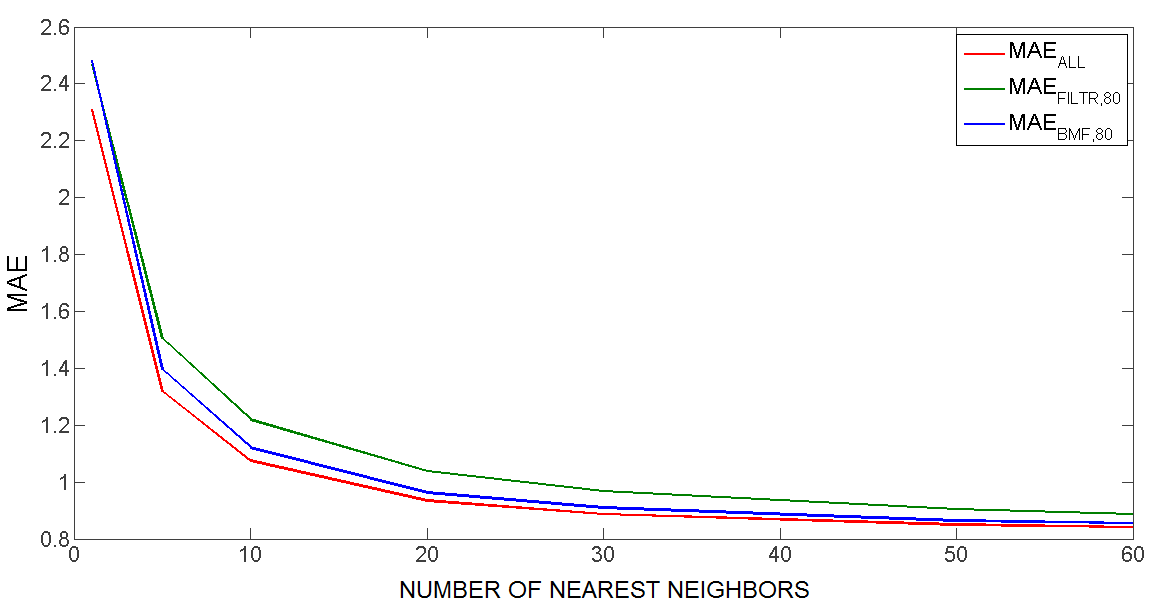}\\
  \caption{\textit{MAE dependance on data   filtration   algorithm    and the  number of nearest  neighbors.}}
  \label{mae_filtr_bmf}
\end{figure}

Just as before to find the nearest neighbors cosine measure is used, and the predicted ratings are calculated as the weighted sum of the ratings of  nearest users.
Figure~\ref{mae_filtr_all} shows that the smaller the data we use for filtering the bigger is  MAE. Figure~\ref{mae_filtr_bmf} shows that the recommendations built on user-factor matrix, are better then  recommendations, constructed on matrix of  ratings  filtered with boolean factorization.

\clearpage

\section{Conclusion}
In the paper we considered main methods of Matrix Factorisation which are suitable for Recommender Systems. Some of these methods were compared on real datasets. We investigated BMF behaviour as part of recommender algorithm. We also conducted several experiments on recommender quality comparison with numeric matrices, user-factor and factor-item matrices in terms of Recall, Precision and MAE. We showed that MAE of our BMF-based approach is not greater than MAE of SVD-based approach for the same number of factors on the real data. For methods that require the number of factors as an initial parameter in the user or item profile (e.g., NMF), we proposed the way of finding this number with FCA-based BMF. We also have investigated how data filtering, namely scaling, influences on recommendations' quality.

As a further research direction we would like to investigate the proposed approaches in case of graded and triadic data \cite{Belohlavek:2012a,Belohlavek:2012b} and reveal whether there are some benefits for the algorithm's quality in usage least-squares data imputation techniques \cite{Wasito:2006}. In the context of matrix factorisation we  also would like to test our approach on the quality assessment of recommender algorithms that we performed on some basic algorithms (see bimodal cross-validation in \cite{Ignatov:2012a}).

\subsubsection{Acknowledgments.} We would like to thank Radim Belohlavek, Vilem Vychodil and Sergei Kuznetsov for their comments, remarks and explicit and implicit help during the paper preparations. We also express our gratitude to Gulnaz Bagautdinova; she did her bachelor studies under the second author supervision on a similar topic and therefore contributed somehow to this paper.

\end{document}